# Resonant excitation of twisted spin waves in magnetic vortices using rotating magnetic fields

Peiyuan Huang[1,2] and Ruifang Wang[1,2*]

[1]*Department of physics, Xiamen University, Xiamen 361005, China*
[2]*Key Laboratory of Low Dimensional Condensed Matter Physics, Department of Education of Fujian Province, Xiamen University, Xiamen 361005, China*

Email: wangrf@xmu.edu.cn

Twisted spin waves attracted research attentions lately and the orbital angular momentum they carry may be utilized in communication and computing technologies. In this work, we manifest by micromagnetic simulations that twisted spin wave modes naturally exist in thick ferromagnetic disks. The twisted spin waves can be readily stimulated using rotating magnetic field when it is tuned to the eigenfrequency of corresponding modes. We analytically derive dispersion relation of the twisted spin waves and the results agree well with the numerical studies. Lastly, we demonstrate that the topological charge of twisted spin waves can be controlled by the exciting field.

Vortex waves carrying orbital angular momentum (OAM) have been discovered in photon,[1,2] electron,[3-5] neutron[6] and phonon[7,8] beams in the past three decades. The phase front of vortex beams twists around its symmetric axis and can be expressed as $\exp(-il\phi)$.[2] The azimuthal angle $\phi$ is defined in the beam's cross section, and $l$ is the quantum number of OAM (also called topological charge), which can be any integer. Most recently, spin waves carrying OAM in ferromagnetic materials were reported. The OAM waves were commonly generated using a Laguerre-Gaussian (LG) excitation field[1,2], diffraction holograms[2,3] or spiral phase plates.[2-4] The topological charge $l$ of twisted waves can be far bigger than one, so can be utilized in multiplex communication and quantum information technologies.[4,6]

In 2019, Chenglong Jia and collaborators[9] discovered twisted spin waves after applying a LG field at one end of a micron sized yttrium iron garnet (YIG) cylinder. Spin waves carrying OAM were also observed in a FeB nanodisk with perpendicular crystalline anisotropy,[10] and in a YIG nano-pillar by passing plane waves through a permalloy ($Ni_{0.8}Fe_{0.2}$) spiral phase plate.[11] In 2020, Peng Yan, et.al. stimulated twisted spin waves with topological charge $l$ as high as $\pm 8$ in a YIG cylinder using LG field, and predicted an exotic "magnetic tweezer" effect of the spin waves carrying OAM.[12] In this paper, we demonstrate that twisted spin wave eigenmodes exist in a thick permalloy nano-disk that was initially in a magnetic vortex state, which is well studied since it was experimentally observed in 2000.[13,14] Spin waves carrying topological charge $l = \pm 1$ can be resonantly excited by a rotating magnetic field, which in experiment can be readily produced at the cross of two perpendicular electric wires. The dispersion relation of the twisted spin waves is derived theoretically, which agrees well with the numerical results. We further show that a spatially modulated rotating field can stimulate twisted spin waves with $l = \pm 2$.

Our model system is a thick permalloy (Py: $Ni_{0.8}Fe_{0.2}$) disk, which diameter and height are both 200 nanometers, as shown in Fig.1(a). Typical material parameters for Py are used in the micromagnetic simulations: saturation magnetization $M_S = 800 \text{ KA}/m$, exchange stiffness constant $A_{ex} = 13 \text{ pJ/m}$, Gilbert damping constant $\alpha = 0.01$ and zero magnetocrystalline anisotropy. The mesh cell size used is $2 \times 2 \times 5$ nm$^3$. All the mircomagnetic simulations employ the OOMMF code. The competition between exchange and magnetostatic interactions leads to a stable vortex state in the sample. The magnetization circulates in-plane around a core, where the magnetization points out-of-plane. Figure 1(a) shows a vortex state that the in-plane magnetization circulates counter clockwise, denoting the vortex chirality $c = 1$. The upward magnetization of the vortex core defines the vortex polarity as $p = +1$.[16,17] To stimulate spin wave oscillation in the waveguide, an in-plane sinc-function field, $\vec{B}(t) = \vec{e}_x B_0 \sin[2\pi f(t - t_0)]/[2\pi f(t - t_0)]$, with $B_0 = 1$mT, $t_0 = 1$ns and the cut-off frequency $f = 20$ GHz, is applied at one end of the sample, i.e., $0 \leq z \leq 5$ nm. We then conduct fast Fourier transformation (FFT) on the subsequent temporal oscillation of $\langle m_x \rangle = \langle M_x \rangle / M_s$, namely the average x-component of magnetization.[18,19] The FFT power spectrum in the frequency domain is shown in Fig.1(b), which contains several resonance peaks. The peak at $f = 0.35$ GHz corresponds to the gyrotropic mode[17] of the vortex. Additional peaks are found at $f$ = 1.6, 3.5, 4.8, 5.9, 7.3, and 8.1 GHz, respectively. After conducting inverse fast Fourier transformation (IFFT)[18] at these eigenfrequencies, we obtain both the FFT amplitude distributions of the corresponding eigenmodes as shown in Fig. 1(c). Clearly, the phase front of the spin wave rotates while the wave propagates along z, an indication of orbital angular momentum.[2,9] It is noteworthy that for vortex states with polarity $p = +1$, only the eigenmodes with topological number $l = -1$ are observed in the cylindrical waveguide. Additional calculations on the identical sample, but with polarity $p = -1$, show that it has the same FFT power spectrum and eigenfrequencies as illustrated in Figure 1(b). However, for this sample, all the eigenmodes have topological number $l = +1$.

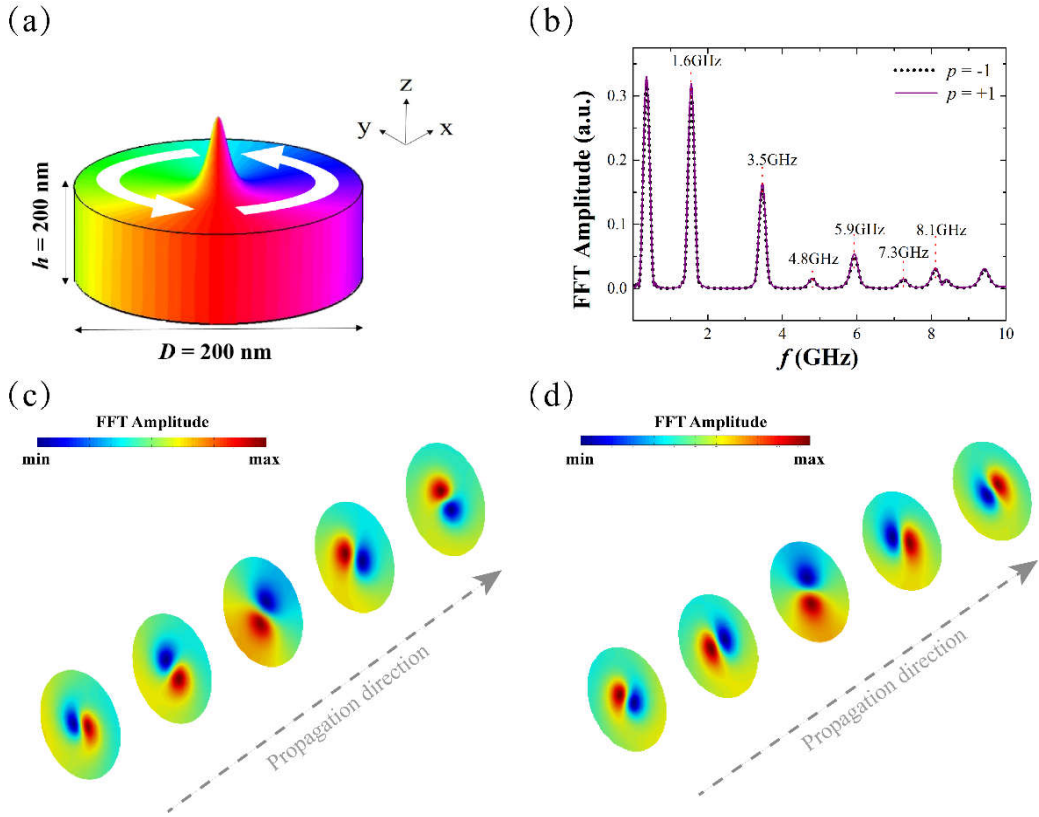

**Fig. 1.** (a) Schematic view of the vortex state in a permalloy cylinder with radius of 100 nm and height of 200 nm. The magnetic vortex has polarity $p = +1$ and chirality $c = +1$. (b) FFT power spectrum of $<m_x>$ after the excitation of an in-plane sinc-function field. Additional calculations on the same sample, but with polarity $p = -1$, deliver the same FFT result. (c) Inverse FFT at $f = 3.5$GHz calculated at different location along the z direction for the vortex spin wave shows topological charge $l=-1$. (d) Additional result, on sample with polarity $p = -1$, of the inverse FFT at $f = 3.5$GHz calculated at different location along the z direction for the vortex spin wave demonstrates topological charge $l=+1$.

After finding out the eigenmodes of vortex spin waves, an in-plane, rotating magnetic field is applied at the $z = 0$ layer of the waveguide. The amplitude of the rotating field is 5 mT and the frequency is set as 3.5 GHz in order to excite the corresponding modes. For the vortex state with polarity $p = +1$, the external field rotates counter clockwise and the excited spin wave has topological charge $l = -1$. But for the sample with polarity $p = -1$, the excitation field must rotate clockwise to generate spin wave with topological charge $l = 1$. Figs. 2(a) and 2(b) show the snapshots of the z-component of the twisted spin waves, 300 ps after application of the rotating field, with topological charges $l = -1$ and $l = 1$, respectively. The corresponding spiral phase fronts are shown in Figs. 2(c) and 2(d), respectively.

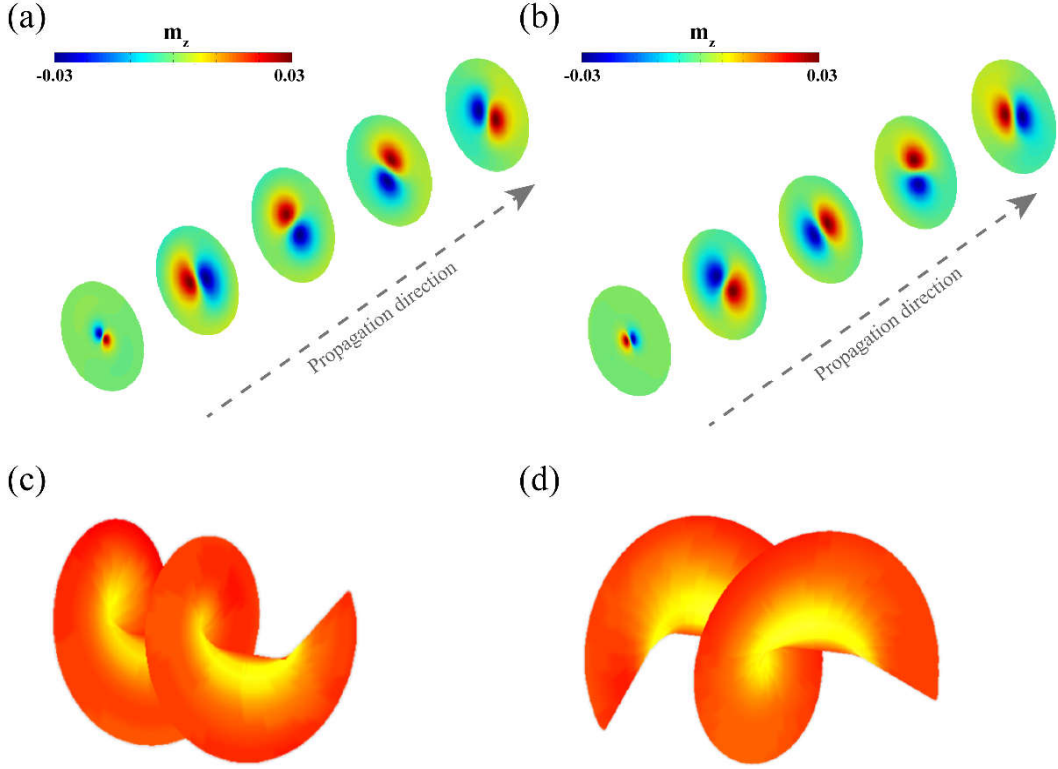

**Fig. 2.** Twisted spin waves propagating along the cylindrical waveguide, which is excited by a 3.5GHz in-plane counter clockwise and clockwise rotating field respectively. Snapshots of the z-component of the twisted spin waves with (a) $l=-1$ and (b) $l = +1$, 0.3 ns after application of external field. The corresponding phase fronts are illustrated in （c）and (d), respectively.

Now, we proceed to derive the dispersion relationship of the vortex spin waves analytically. The magnetization dynamics is described by the Landau-Lifshitz-Gilbert (LLG) equation,

$$\frac{\partial \bm{m}}{\partial t} = -\gamma \mu_0 \bm{m} \times \bm{H}_{eff} + \alpha \bm{m} \times \frac{\partial \bm{m}}{\partial t} \quad (1)$$

with the gyromagnetic ratio $\gamma = 1.76 \times 10^{11}\,\text{rad}\,\text{s}^{-1}\,\text{T}^{-1}$, the vacuum permeability $\mu_0$, and the Gilbert damping $\alpha$. After neglecting the second term on the right side of LLG equation, since $\alpha$ is only 0.01 for the Py sample, the LLG equation is reduced as,

$$\partial \bm{m}/\partial t = -\gamma \mu_0 \bm{m} \times \bm{H}_{eff} \quad (2)$$

Using cylindrical coordinates $(\rho, \phi, z)$, the out-of-plane magnetization component of the vortex spin wave can be approximately expressed as,[9]

$$\Delta m_z(\rho, \phi, z, t) = J_r(k_\perp \rho) exp(il\phi + ik_z z) exp(-i\omega t) \quad (3)$$

where $J_r(x)$ is the Bessel function of the first kind with order $r$. Here we consider the case of $r = 1$, thus equation (3) becomes

$$\Delta m_z(\rho, \phi, z, t) = J_1(k_\perp r) exp(i\phi + ik_z z) exp(-i\omega t) \quad (4)$$

For simplicity, we neglect the contribution of dipolar field and consider only the contribution of exchange field in the effective field $\mathbf{H}_{eff}$. Thus,

$$\gamma \mathbf{H}_{eff} = \tilde{A} \nabla^2 \mathbf{m} \quad (5)$$

where, $\tilde{A} = 2\gamma A_{ex}/M$. After substituting Eqs. (4) and (5) into (2), the spin wave dispersion relationship is obtained as,

$$\omega(k_z) = \tilde{A}(k_z^2 + k_\perp^2) \quad (6)$$

where $k_z = 2\pi/\lambda$ and $k_\perp = x_0/R$, in which $\lambda$ is defined as the wavelength, i.e., the wave propagation length after the phase changes $2\pi$, $x_0$ is the first zero point of the first-order Bessel function, and $R$ is the radius of the cylindrical waveguide.

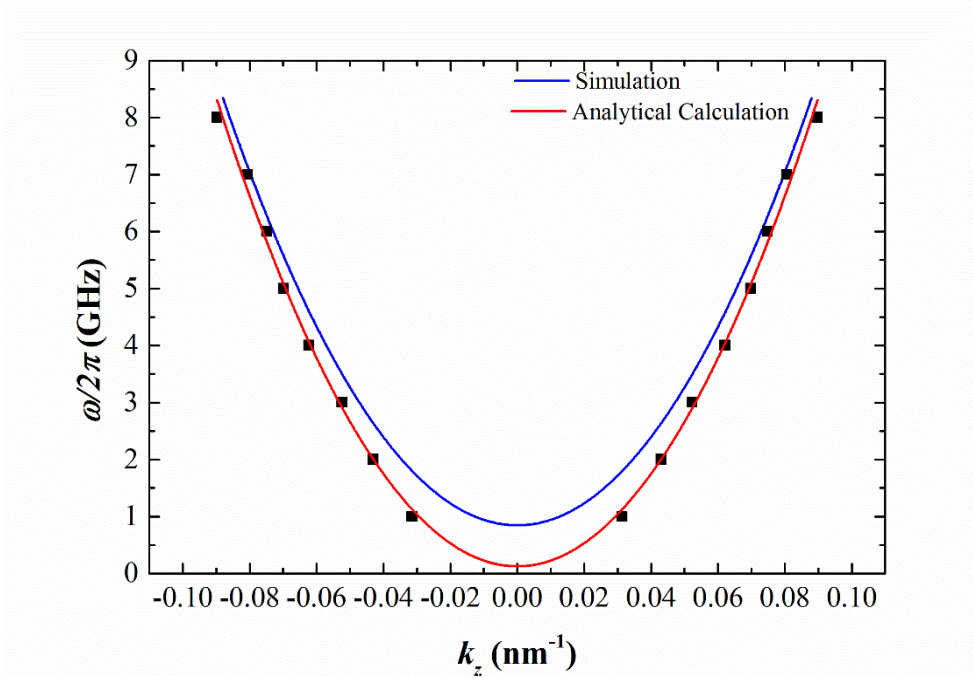

**Fig.3.** The dispersion relationship of twisted spin waves with $l = +1$. The black squares represent numerical simulation results and the red solid line is the fitting curve. The solid blue line is obtained from analytical calculation.

Figure 3 illustrates that the result of Eq. (6) agrees well with the numerical simulations for spin wave frequency above 5 GHz, because the exchange energy dominates in this range and the dipolar interaction can be neglected,[20] which is the assumption when the analytical result is derived. As the wave frequency decreases, the discrepancy between theory and the numerical result grows, resulting from growing importance of magnetic dipole interaction.

To generate vortex spin waves with topological charges $l = +2$ or $l = -2$, a spatially non-uniform rotating field is applied. At $t = 0$, the excitation field orients along $\hat{x}, -\hat{y}, -\hat{x}$ and $\hat{y}$ in four square regions as shown in the Fig. 4(a). Each region is 100 nm by 100 nm. The excitation field locates at the $z = 0$ layer and rotates in the x-y plane, with oscillation amplitude of 5 mT and frequency of 7 GHz. The topological charge of stimulated spin wave depends on the rotating sense of excitation field. A counter clockwise field stimulate waves with $l = -2$, and clockwise field generates waves with $l = +2$. Figs. 4(c) and 4(d) show snapshot of the z-component magnetization excitation propagating after 0.3 ns. This result demonstrate that the topological charges of vortex spin waves can be manipulated by the topology of external field.

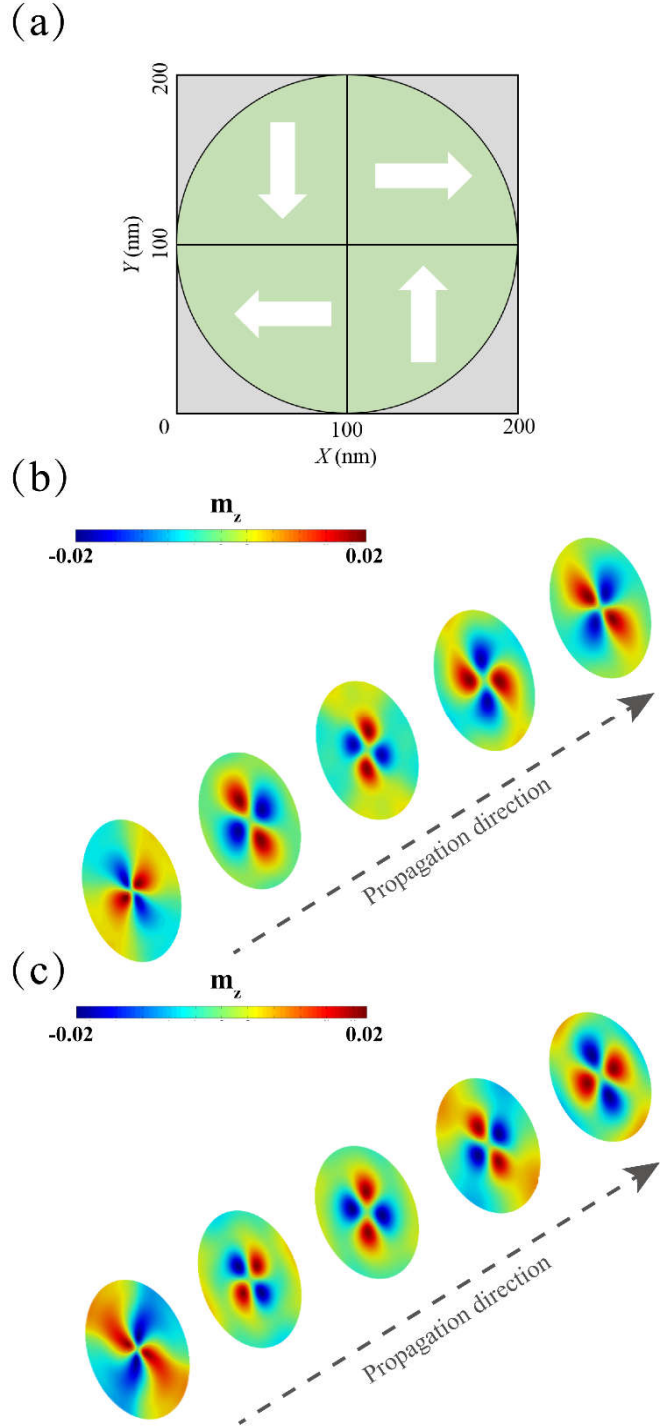

**Fig. 4.** (a) The orientation, shown as arrows, of the spatially non-uniform excitation field at $t = 0$. Snapshots of the z-component of the twisted spin waves with (b) $l=-2$ and (c) $l = +2$, 0.3 ns after application of external field.

In summary, our micromagnetic simulations manifest that twisted spin wave modes exist in a thick permalloy cylinder, which is initially in a vortex state. Spin waves with topological charges $l = \pm 1$ are generated by applying counter clockwise or clockwise rotating magnetic field at the eigenfrequencies. The dispersion relationship of the twisted spin waves is calculated analytically. The theoretical results agree well with numerical calculations for wave frequency higher than 5 GHz. By applying a spatially non-uniform rotating magnetic field, we observe twisted spin waves with topological charges $l = \pm 2$ in the sample.

We acknowledge the financial support from the National Natural Science Foundation of China under grant number 11174238.